\documentclass[prb,showpacs,twocolumn,aps]{revtex4}
\usepackage{graphicx}
\usepackage{color}
\usepackage{dcolumn}
\usepackage{amsmath}
\usepackage{amssymb}

\newcommand{\lsco}{$\rm La_{2- \it x}Sr_{\it x}CuO_{4}$}
\newcommand{\lbco}{$\rm La_{2- \it x}Ba_{\it x}CuO_{4}$}

\newcommand{\lbcoe}{$\rm    La_{1.875}     Ba_{0.125}               CuO_{4}$}
\newcommand{\lbcon}{$\rm    La_{1.905}     Ba_{0.095}               CuO_4$}

\newcommand{\lnescoe}{$\rm  La_{1.88- \it y}$ $\rm (Nd,Eu)_{\it y}$ $\rm Sr_{0.12}  CuO_{4}$}

\newcommand{\ybco}{$\rm YBa_2Cu_3O_{6+\delta}$}

\newcommand{\pla}{$\rm CuO_2$}
\newcommand{\oct}{$\rm CuO_6$}

\include{version}

\begin{document}

\title{Enhanced charge stripe order of superconducting $\bf La_{\it 2-x}Ba_{\it x}CuO_{\it 4}$ in a magnetic field.}
\author{M. H\"ucker$^1$}
\author{M. v. Zimmermann$^2$}
\author{Z. J. Xu$^1$}
\author{J. S. Wen$^1$}
\author{G. D. Gu$^1$}
\author{J. M. Tranquada$^1$}

\affiliation{$^{1}$Condensed Matter Physics \&\ Materials Science Department,
Brookhaven National Laboratory, Upton, New York 11973, USA}
\affiliation{$^{2}$Hamburger Synchrotronstrahlungslabor HASYLAB at Deutsches
Elektronen-Synchrotron DESY, 22603 Hamburg, Germany}

\date{\today}

\begin{abstract}

The effect of a magnetic field on the charge stripe order in \lbco\ has been studied
by means of high energy (100 keV) x-ray diffraction for charge carrier
concentrations ranging from strongly underdoped to optimally doped. We find that
charge stripe order can be significantly enhanced by a magnetic field applied along
the $c$-axis, but only at temperatures and dopings where it coexists with bulk
superconductivity at zero field. The field also increases stripe correlations
between the planes, which can result in an enhanced frustration of the interlayer
Josephson coupling. Close to the famous $x=1/8$ compound, where zero field stripe
order is pronounced and bulk superconductivity is suppressed, charge stripe order is
independent of a magnetic field. The results imply that static stripe order and
three-dimensionally coherent superconductivity are competing ground states.
\end{abstract}


\maketitle

There is mounting evidence for proximity of the superconducting (SC) ground state in the
cuprates to competing states with static spin and/or charge density
modulations.~\cite{Lake02a,Hoffman02a,Hanaguri04a,Fujita04a,Suchaneck10a,wu11} A very
interesting example was recently observed with soft and hard x-ray diffraction in \ybco
-based cuprates.~\cite{Ghiringhelli12a,chan12a} Around a hole concentration in the \pla\
planes of $p=1/8$, both techniques detect the onset of an incommensurate charge density
modulation at $\sim$140~K that decreases below the SC transition at $T_c\sim 65$~K, but
can be enhanced if SC is weakened by a magnetic field applied perpendicular to the \pla\
planes ($H\parallel c$). One much-discussed possibility is that the order is caused by a
nesting instability associated with a reconstruction of the Fermi surface, for which
there is evidence from quantum oscillation
measurements.~\cite{DoironLeyraud07a,LeBoeuf07a, Sebastian07a} \ybco\ also exhibits
incommensurate spin correlations~\cite{Hinkov08a,Haug09a,Haug10a}; however, the magnetic
wave vectors seem to be unrelated to those of the charge modulations. This conclusion is
corroborated by the fact that spin excitations are gapped for $p \gtrsim 0.08$, which
includes the region showing the charge modulations and quantum
oscillations.~\cite{Dai01a,Hayden04a}

Another competing state is the stripe phase in the La-based cuprates which also is
most stable at a hole content of $p\sim 1/8$, where $p=x$.~\cite{Tranquada95a}
Famous examples are \lbcoe\ and \lnescoe\ where bulk SC is strongly suppressed and
replaced by an incommensurate order that has been described as an arrangement of
charge stripes (or charge order, CO) separating antiferromagnetic spin stripes
(spin order, SO).~\cite{Tranquada95a,Fujita04a, Huecker07b,Fink09a} The spin correlations
resemble those in \ybco\ at lower doping~\cite{Haug10a}; however, the CO
wave vector is uniquely related to the SO wave vector.~\cite{Tranquada97a,
Fink11a,Huecker11a} Does this mean that the charge modulations in the Y- and
La-based cuprates have different origins? Understanding their physics seems crucial
and may provide important clues about the SC itself.

To make progress on the stripes frontier, recent studies have focussed on \lbco\ in
high magnetic fields.~\cite{Kim08b,Huecker11b,Wen12b,Stegen12a} If SO and CO are
indeed coupled and compete with SC, both stripe orders should increase by similar
amounts in a magnetic field $H \parallel c$.  The first clear evidence that this is
indeed the case, was obtained in strongly underdoped \lbcon , which is a bulk SC
with weak zero-field stripe order.~\cite{Huecker11b,Wen12b} This observation makes
\lbco\ an excellent system in which to study the field effect on the CO as a function of
doping.

Here we report x-ray diffraction experiments on \lbco\ for $0.095 \leq x \leq 0.155$
and fields up to 10~T. We show that CO can be enhanced in a broad range of doping.
The effect is particularly large in samples far away from $x = 1/8$ where CO is weak
and bulk SC strong, and is absent close to $x=1/8$ where CO is strong and bulk SC
suppressed. It is observed only below $T_c$ and for $H\parallel c$, which implies
that stripe order emerges as the new ground state when bulk SC is suppressed. For
the compositions showing the strongest effect, $x=0.095$ and $0.155$, even at
$H=10$~T the CO order parameter remains much below that at $x=1/8$. We have also
analyzed the CO correlations between the planes, and for $x=0.095$ we find a clear
enhancement due to the field.

\begin{figure*}[t]
\center{\includegraphics[width=2.09\columnwidth,angle=0,clip]{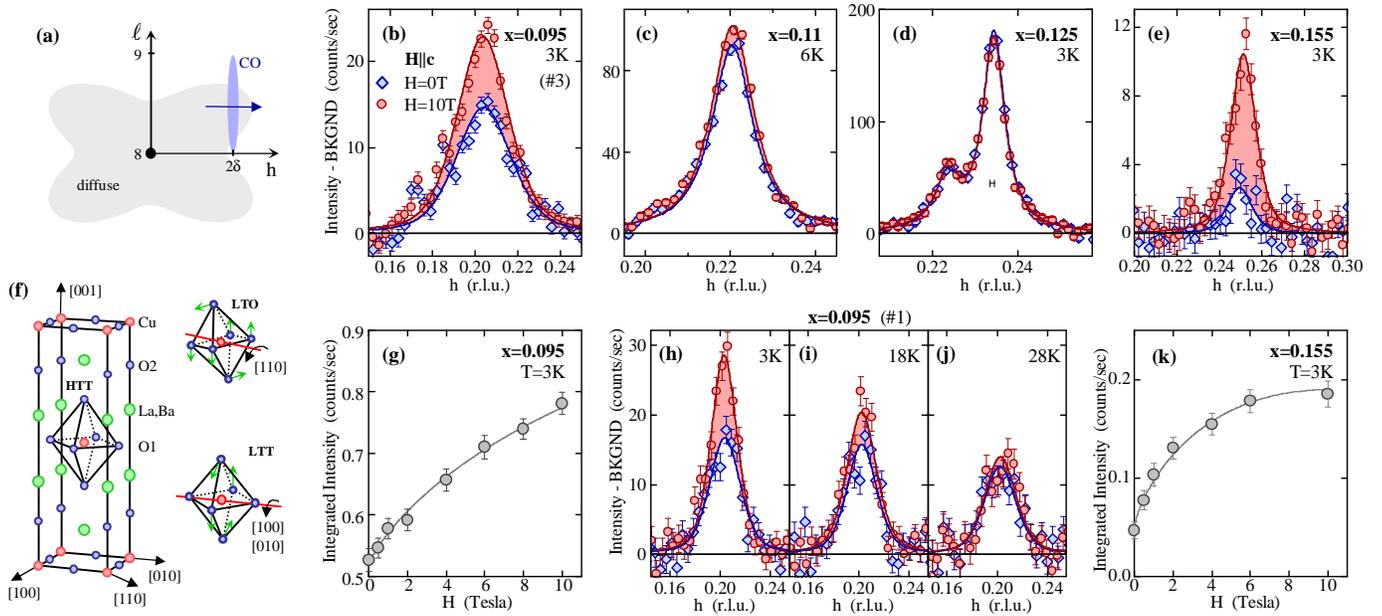}}
\caption[]{(color online) In-plane CO correlations at $T \sim 3$~K for $0.095 \leq x \leq
0.155$ and $H \parallel c$. (a) $(h,0,\ell)$-zone with CO-peak at (2$\delta$, 0, 8.5),
diffuse intensity around (0, 0, 8), and typical $h$-scan. (f) Unit cell in the HTT phase.
Tilt directions of the CuO$_6$ octahedra in the LTO and LTT phases. (b-e) $h$-scans at
$H=0$~T and 10~T for different $x$. (h-j) $h$-scans for $x=0.095$ at different
temperatures. (g,k) Integrated intensity for $x=0.095$ and 0.155. The solid lines are
least squares fits using $I(H)=I_0+I_1(H/H_{c2})\ln(H_{c2}/H)$ of
Ref.~\onlinecite{Demler01a}, where the upper critical field of the SC state $H_{c2}$, and
$I_0$ and $I_1$ are parameters. In agreement with expectations we find that $H_{c2}$ is
larger for $x=0.095$ than for $x=0.155$. (d) The split of the CO-peak for $x=0.125$ is
caused by the crystal's mosaic.~\cite{Huecker11a} The horizontal bar indicates the
instrumental resolution full width at half maximum. Solid lines through the $h$-scans are
least squares fits using a pseudo-Voigt line shape. }\label{fig1}
\end{figure*}

The \lbco\ single crystals with nominal Ba contents $x=0.095$, 0.11, 0.115, 0.125, 0.135,
and 0.155 are the same as in our zero-field study\cite{Huecker11a}; some of these
compositions have been the subject of further characterizations.\cite{Tranquada08a,
Wilkins11a,Homes12a,Wen12a,Wen12b,Stegen12a,li11,karapetyan12}  Figure~\ref{fig1}(f)
shows the crystal structure of \lbco , which differs from that of \lsco\ in a subtle
fashion that explains their distinct behaviors.~\cite{Axe89,Buechner94c,huck12} At low
temperature \lsco\ assumes orthorhombic (LTO) symmetry ($Bmab$), whereas \lbco\
transforms from LTO to tetragonal (LTT) symmetry ($P4_2/ncm$), or a less orthorhombic
(LTLO) symmetry ($Pccn$), which is a structure between LTO and
LTT.~\cite{Huecker11a,Wen12b} In the LTO phase the \oct\ octahedra tilt about [110],
causing all in-plane Cu-O-Cu bonds to bend, whereas in the LTT phase they alternately
tilt about [100] and [010] in adjacent planes, causing only half of all bonds to bend.
This locally broken rotational symmetry of the \pla\ planes in the LTT phase is believed
to pin stripes more strongly.~\cite{Tranquada95a}

The x-ray diffraction experiments were performed with the triple-axis diffractometer
at beamline BW5 at DESY at a photon energy of 100~keV.~\cite{Bouchard98} The
crystals were mounted with the $(h,0,\ell)$-zone in the scattering plane, and the
magnetic field was applied parallel to the $c$-axes. Further experimental details
have been described in Ref.~\onlinecite{Huecker11a}. Scattering vectors ${\bf
Q}=(h,k,\ell)$ are specified in units of $(2\pi/a, 2\pi/a, 2\pi/c)$, where
$a=3.78$~\AA\ and $c=13.2$~\AA\ are the lattice parameters of the high temperature
tetragonal (HTT) phase ($I4/mmm$) in Fig.~\ref{fig1}(f). The data for $x=0.095$ was
obtained in three experiments, which we indicate by numbers (\#) in the figures.
\begin{figure*}[t]
\center{\includegraphics[width=2.09\columnwidth,angle=0,clip]{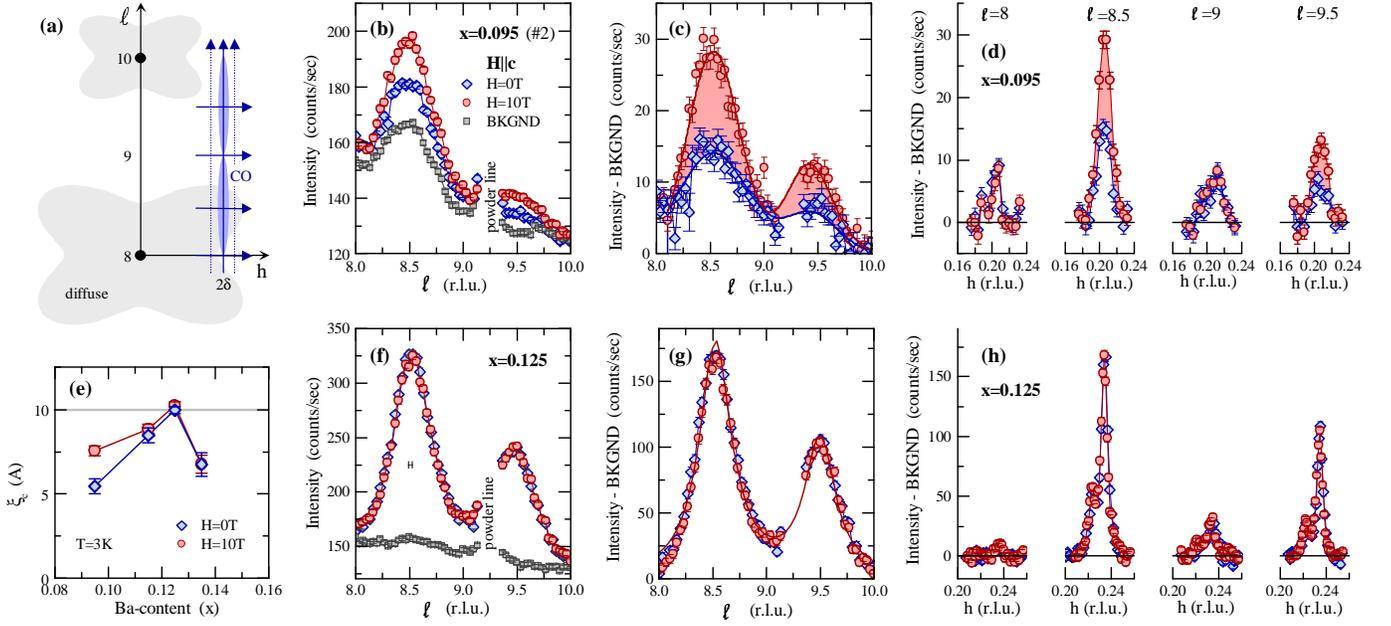}}
\caption[]{(color online) Out-of-plane CO correlations at $T=3$~K for $x=0.095$ and 0.155
vs field $H\parallel c$. (a) $(h,0,\ell)$-zone with CO-peaks at $\ell = 8.5$ and 9.5,
diffuse intensity around fundamental peaks, and typical $\ell$ and $h$-scans. (b,f)
$\ell$-scans at $H=0$~T and 10~T, including background from average of $\ell$-scans at
$h=2\delta \pm 0.03$. The horizontal bar in (f) indicates the instrumental resolution
full width at half maximum. (c,g) Same data after background subtraction, and with least
squares fits using a pseudo-Voigt line shape. (d,h) $h$-scans at various $\ell$-values.
(e) Correlation length $\xi_c$ vs $x$.}\label{fig2}
\end{figure*}

The CO leads to weak satellites about the fundamental reflections with ordering wave
vectors ${\bf Q}_{\rm CO} = (2 \delta, 0, 0.5)$ and $(0, 2\delta, 0.5)$. To study the CO
within the \pla\ planes we have performed $h$-scans through the satellite at $(2\delta,
0, 8.5)$, indicated in Fig.~\ref{fig1}(a). Figure~\ref{fig1}(b-e) displays data at $H=0$
and 10~T applied $\parallel c$ for four dopings at base temperature. Obviously, 10~T
results in large intensity gains for $x<1/8$ and $x>1/8$, but does not affect peak
positions. In particular, for $x=0.095$ the CO-peak increases by $\sim$50\%, and for
$x=0.155$ by $\sim$200\%. The field effect decreases toward zero as $x \rightarrow 1/8$,
as is shown in Fig.~\ref{fig1}(c,d). Already at $x=0.11$ the enhancement is very small
($< 10\%$). But it was confirmed to be finite in a second experiment. Additional data
(not shown) for $x=0.115$ and 0.135 show no effect.

The detailed field dependence of the integrated intensity for $x=0.095$ and $x=0.155$ in
Fig.~\ref{fig1}(g,k) shows a strong initial increase followed by a tendency to saturate,
which is similar to the SO in \lsco .~\cite{Lake02a}  In Ref.~\onlinecite{Demler01a} it
was predicted for a state of coexisting spin-density-wave and superconducting order that
the SO intensity should grow as $(H/H_{c2})\ln(H_{c2}/H)$, a form that is consistent with
experimental studies.\cite{Lake02a,khay02,Chang08a}  We find that the same functional
form describes the CO data; we emphasize, however, that the model in
Ref.~\onlinecite{Demler01a} did not explicitly include any charge stripe order.

To determine whether the field affects the stripe stacking order along the $c$-axis, we
performed $\ell$-scans through the CO-peaks at $(2 \delta, 0, 8.5)$ and $(2 \delta, 0,
9.5)$, as is indicated in Fig.~\ref{fig2}(a). Again $x=0.095$, in Fig.~\ref{fig2}(b),
shows a strong field effect while $x=0.125$, in Fig.~\ref{fig2}(f), is constant.
Additional $\ell$-scans at $h=2\delta \pm 0.03$ were performed to estimate the background
signal at $h=2\delta$. This is particularly important for $x=0.095$ where the CO-peak is
small and the background has a similar $\ell$-dependence due to a contribution from
diffuse scattering around the Bragg peaks. The corrected data in Fig.~\ref{fig2}(c,g)
were fit to extract the peak widths, and from that the $c$-axis correlation lengths
$\xi_c$ in Fig.~\ref{fig2}(e). While $\xi_c \sim 10$~\AA\ for $x=0.125$, which is
slightly below one lattice constant, it is only half of that for $x=0.095$ at zero field,
but it is enhanced by 50\%\ at 10~T.

To confirm this result for $x=0.095$, $h$-scans at different $\ell$ were performed,
see Fig.~\ref{fig2}(d,h). If the field were to increase only the peak intensity,
percentagewise it should be the same at any $\ell$. This is clearly not the case.
Intensity increases at the peak positions $\ell = 8.5$ and 9.5, but not at $\ell =
9$ where the tails of the peaks overlap, which implies that the peaks indeed narrow
in $\ell$ and that $\xi_c$ grows.

\begin{figure}[!t]
\center{\includegraphics[width=0.9\columnwidth,angle=0,clip]{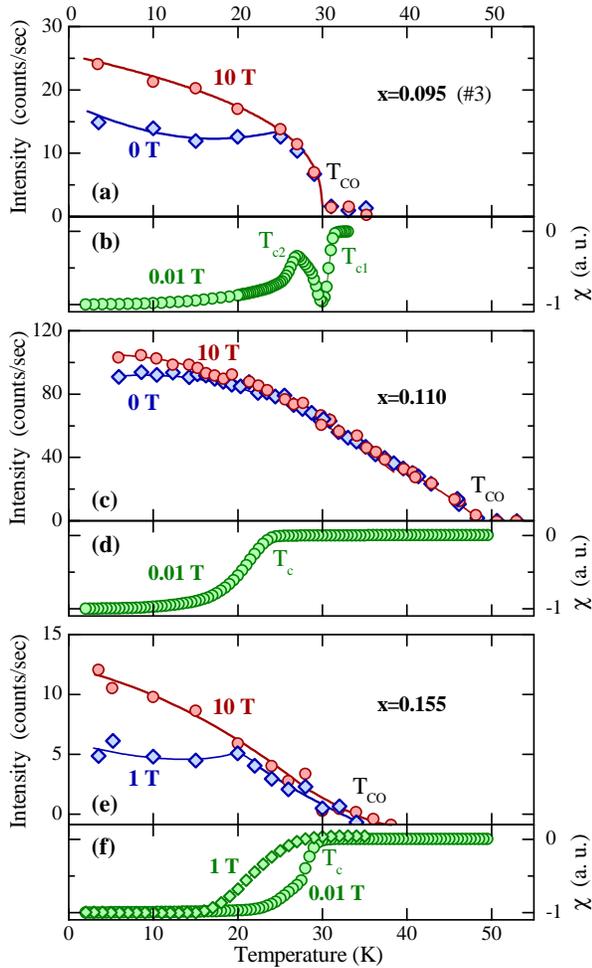}}
\caption[]{(color online) $T$-dependence of CO and SC for $x=0.095$, 0.11, and 0.155 and
$H\parallel c$. (a,c,e) Peak intensity of (2$\delta$, 0, 8.5) CO-peak at $H=0$~T (1~T for
$x=0.155$) and 10~T. (b,d,f) Normalized Meissner effect at $H=0.01$~T.}\label{fig3}
\end{figure}

Next we look at the $T$-dependence. Representative $h$-scans for $x=0.095$ in
Fig.~\ref{fig1}(h-j) indicate that the field effect is maximum at low $T$ and disappears
upon warming. Figure~\ref{fig3} presents more detailed data for the three dopings that
show a field effect. The top panels display peak intensities at 0~T (1~T for x=0.155) and
10~T, the lower panels the Meissner effect at 0.01 T. For all three dopings the CO
depends on the field only below $T_c$. The $x=0.095$ crystal displays a particularly
interesting SC transition that is interrupted by the CO
transition.~\cite{Huecker11a,Huecker11b,Wen12a,Wen12b} SC first appears at $T_{\rm
c1}=32$~K in the non-stripe-ordered LTO phase. This SC state weakens when CO sets in at
the LTO$\rightarrow$LTLO transition at 30~K, with a corresponding reduction of the
interlayer Josephson coupling.\cite{Homes12a}  (Note that $x=0.095$ is nearly
LTT.~\cite{Huecker11a,Wen12b}) Then at $T_{\rm c2}=27$~K SC reappears, at which point the
zero field CO saturates. Only when suppressing the SC state with 10~T, the CO-peak
continuous to increase below $T_{\rm c2}$. Note that the $T$-dependence in
Fig.~\ref{fig3}(a) was measured with higher precision than in Ref.~\onlinecite{Wen12b},
and now reveals the impact of SC on the CO below $T_{\rm c2}$. Also in the case of
$x=0.155$, where the CO was measured with a minimum field of 1~T, the onset of the field
effect coincides with the SC transition measured in the same field.

Could all these effects be the result of a magneto-elastic mechanism that enhances stripe
pinning? The most relevant pinning parameter of the LTT and LTLO phases is the \oct\ tilt
angle $\Phi$.~\cite{Buechner94c} If $\Phi$ were to increase with field certain
superstructure reflections, such as (1,~0,~0), would become stronger. We find these peaks
to be independent of $H$, which leads us to conclude that the CO enhancement is a purely
electronic effect.

\begin{figure}[!t]
\center{\includegraphics[width=1\columnwidth,angle=0,clip]{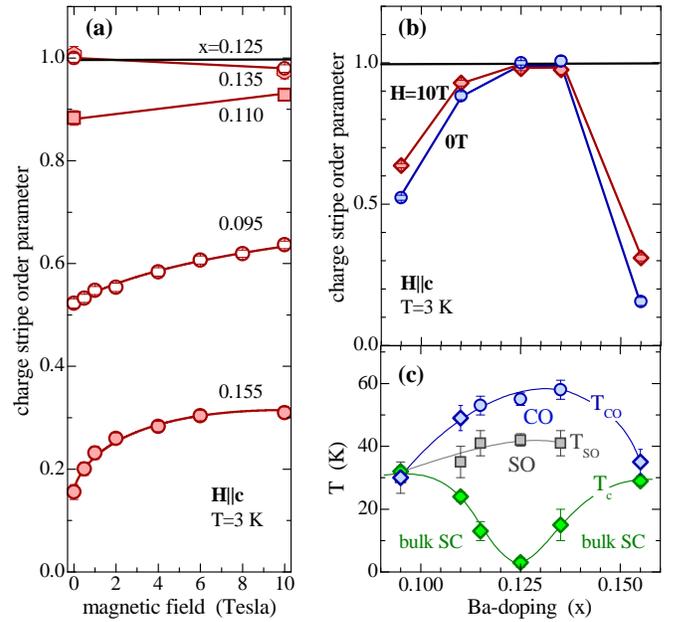}}
\caption[]{(color online) CO order parameter for $ H \parallel c$ and different $x$ at
$T=3$~K. (a) As a function of $H$. (b) As a function of doping at $H=0$~T and 10~T. (c)
Zero field phase diagram with critical temperatures $T_c$, $T_{\rm CO}$, and $T_{\rm SO}$
from Ref.~\onlinecite{Huecker11a}, except for three new $T_{\rm CO}$ values (diamonds)
from present study. The solid lines for $x=0.095$ and 0.155 in (a) are fits using the
square root of the expression in Fig.~\ref{fig1}. All other solid lines are guides to the
eye.}\label{fig4}
\end{figure}

In Fig.~\ref{fig4}(a) we plot the doping and field dependence of the CO order parameter,
normalized to $x=1/8$ in zero field. The data represent the square root of the integrated
intensity of the $h$-scans. Strongly underdoped $x=0.095$ and optimally doped $x=0.155$
display a strong initial increase, but tend to saturate at high fields at $\sim$65\% and
$\sim$30\% of the full order at $x=1/8$, respectively. The crystals closer to $x=1/8$
show order parameters larger than $\sim$90\% and either no or just a weak increase with
field. (The $x=0.115$ sample also shows no field effect, but that crystal was measured
under different conditions which impedes a direct comparison.) In Fig.~\ref{fig4}(b,c) we
compare the doping dependence of the CO order parameter at $H=0$~T and 10~T with that of
$T_c$, $T_{\rm CO}$, and $T_{\rm SO}$ in zero field.~\cite{Huecker11a} Clearly, the CO
enhancement is maximum where bulk SC is strong and CO is weak. This corresponds well with
the neutron diffraction data for $x=0.095$ and 0.125 which show a similar $H$-dependence
of the SO.~\cite{Wen12a,Wen08a} The weak enhancement of the SO close to $T_{\rm SO}$
reported in Ref.~\onlinecite{Wen08a} for $x=0.125$ in high fields, is not observed for
the CO close to either $T_{\rm SO}$ or $T_{\rm CO}$. We assume that the SO is stabilized
not only through the suppression of SC, but also through the gain of Zeeman energy.

We note that the observed field effect could also represent a change of the stripe
ordered volume fraction proportional to that of the integrated intensity.  For example,
if the stripe order is induced in the vicinity of magnetic vortices,\cite{kive02} then
the intensity might grow with the vortex density (proportional to $H$) until the CO
correlation length\cite{Huecker11a} becomes comparable to half of the vortex spacing,
which occurs near 10~T for $x=0.095$ and $x=0.155$.  Of course, any evaluation of volume
fractions would depend on the local maximum order parameter, which we do not have
independent knowledge of.

The enhancement of CO at high magnetic fields in \lbco\ over a broad range of $x$ is a
long-sought-for confirmation of the strong coupling between stripe type charge and spin
orders in La-based cuprates. In the case of \lsco\ several neutron scattering experiments
of the past decade have shown that SO can be enhanced by a field $H\parallel
c$.~\cite{Lake02a,Chang08a,Chang09a} However, there had been no evidence of CO until very
recent resonant soft x-ray scattering experiments revealed CO near the sample surface,
but not in the bulk.~\cite{Wu12a} Our high-energy x-ray data prove that in \lbco\ the
zero field CO and its enhancement in the field are bulk properties. The exclusive
occurrence of the field effect in bulk SC below $T_c$, as well as on both sides of
$x=1/8$ doping, clearly implies a competition between stripe order and SC.

Recently it was proposed that stripe order does not suppress SC pairing correlations in
the planes, but prevents three dimensional phase coherence by frustrating the interlayer
Josephson coupling.~\cite{Berg07a,Berg09a,Li07c,Tranquada08a,Wen12b} Thus, it is possible
that the field not only suppresses SC, but also enhances the interlayer CO correlations.
The field driven increase of the CO correlation length $\xi_c$ for $x=0.095$ is clear
evidence of such an effect.

In \lbco\ the zero field CO wave vector is tightly linked to that of the SO, and
increases with $x$, in agreement with the trend predicted by the stripe
model.~\cite{Tranquada95a,Huecker11a,Fujita12a} Our study shows that this trend is
independent of the magnetic field. Furthermore, the increase of the CO wave vector is
incompatible with the decrease of the antinodal nesting vector, as measured with
angle-resolved photo-emission spectroscopy.~\cite{Valla06a} This is different for the
checkerboard type charge modulation in the Bi-based cuprates, and the recently discovered
modulations in Y-based compounds.~\cite{Hoffman02a,Wise08a,achkar12,chan12a} There the
charge modulation wave vectors either decrease with doping or stay approximately
constant, and tend to agree with a Fermi surface nesting
scenario.~\cite{Wise08a,achkar12,Ghiringhelli12a,chan12a} Thus, the sum of experiments
seems to indicate a distinct nature for the stripe order in La-based cuprates, and the
nesting related charge modulations in Bi- and Y-based cuprates. However, the
qualitatively same field dependence of these two states in the normal state as well as
below $T_c$, as observed here and in Ref.~\onlinecite{chan12a}, suggests that they
depend in a similar way on the suppression of the competing bulk SC state. This makes one
wonder if and how these charge modulated states are connected, which is the next piece of
the cuprate puzzle to understand.

The work at Brookhaven was supported by the Office of Basic Energy Sciences, Division of
Materials Science and Engineering, U.S. Department of Energy (DOE), under Contract No.
DE-AC02-98CH10886.

%\bibliographystyle{apsrev}
%\bibliography{mh,more}
%\begin{thebibliography}{10}
%\end{thebibliography}

\end{document}